\documentclass[letterpaper]{article} % DO NOT CHANGE THIS
\usepackage{aaai25}  % DO NOT CHANGE THIS
\usepackage{times}  % DO NOT CHANGE THIS
\usepackage{helvet}  % DO NOT CHANGE THIS
\usepackage{courier}  % DO NOT CHANGE THIS
\usepackage[hyphens]{url}  % DO NOT CHANGE THIS
\usepackage{graphicx} % DO NOT CHANGE THIS
\usepackage{natbib}  % DO NOT CHANGE THIS AND DO NOT ADD ANY OPTIONS TO IT
\usepackage{caption} % DO NOT CHANGE THIS AND DO NOT ADD ANY OPTIONS TO IT
\usepackage{amsmath,amssymb}
\newcommand{\parsection}[1]{\noindent\textbf{#1}:}

\title{Optimizing Gene-Based Testing for Antibiotic Resistance Prediction}
\author {
    David Hagerman\textsuperscript{\rm 1},
    Anna Johnning\textsuperscript{\rm 1,\rm 2},
    Roman Naeem\textsuperscript{\rm 1}, \\
    Fredrik Kahl\textsuperscript{\rm 1},
    Erik Kristiansson\textsuperscript{\rm 1,\rm 2},
    Lennart Svensson\textsuperscript{\rm 1}
}
\affiliations {
    \textsuperscript{\rm 1}Chalmers University of Technology \\
    \textsuperscript{\rm 2}Fraunhofer-Chalmers Centre \\
    \{david.hagerman, anna.johnning, nroman\}@chalmers.se, 
    \\
    \{fredrik.kahl, 
    erik.kristiansson, 
    lennart.svensson\}@chalmers.se
}

\begin{document}

\maketitle

\begin{abstract}
Antibiotic Resistance (AR) is a critical global health challenge that necessitates the development of cost-effective, efficient, and accurate diagnostic tools. Given the genetic basis of AR, techniques such as Polymerase Chain Reaction (PCR) that target specific resistance genes offer a promising approach for predictive diagnostics using a limited set of key genes. This study introduces GenoARM, a novel framework that integrates reinforcement learning (RL) with transformer-based models to optimize the selection of PCR gene tests and improve AR predictions, leveraging observed metadata for improved accuracy. 
In our evaluation, we developed several high-performing baselines and compared them using publicly available datasets derived from real-world bacterial samples representing multiple clinically relevant pathogens. The results show that all evaluated methods achieve strong and reliable performance when metadata is not utilized. When metadata is introduced and the number of selected genes increases, GenoARM demonstrates superior performance due to its capacity to approximate rewards for unseen and sparse combinations. Overall, our framework represents a major advancement in optimizing diagnostic tools for AR in clinical settings.
\end{abstract}

\begin{links}
    \link{Datasets}{https://www.ncbi.nlm.nih.gov/pathogens/}
    \link{Code}{https://github.com/Eiphodos/GenoPhen}
\end{links}

\section{Introduction}

Antimicrobial Resistance (AMR) poses an escalating challenge to global health, estimated to result in more than 10 million yearly deaths by 2050~\cite{laxminarayan2013antibiotic, o2014antimicrobial, o2016tackling}, the overwhelming majority of which are attributed to antibiotic resistance (AR). Pathogens become resistant to antibiotics through mutations in their genome, either by mutations in pre-existing genes or by acquiring antibiotic-resistance genes from other bacterial cells. The rate at which pathogens acquire resistance traits significantly exceeds the pace of new drug development, underscoring the urgent need for increased research efforts in this domain. As a result, there is a strong demand for tools that assist in prescribing the most effective antibiotics against infections. These tools should be capable of estimating antibiotic resistance with high precision, enabling rapid identification of resistant strains and targeted treatment, which could reduce antibiotic misuse and slow the spread of resistance.

Recent studies \cite{tharmakulasingam_explainable_2022, tharmakulasingam_rectified_2023, macesic_predicting_2020, jin_predicting_2024, kuang_accurate_2022} have shown that deep learning models using complete pathogen genomic information can accurately predict AR. However, the requirement for full genome sequencing presents significant obstacles in a clinical context, primarily due to its high cost, slow processing times, and overall impracticality for rapid decision-making. An alternative approach involves using tests based on Polymerase Chain Reaction (PCR) amplification, which focuses on analyzing a targeted subset of known resistance mutations and/or genes. Compared to full gene sequencing, PCR tests are faster, more cost-effective, and therefore often more suitable for clinical use.

While PCR tests offer a faster and cheaper alternative to full gene sequencing, they require the careful selection of a specific subset of genes for testing. With more than 5,000 genes and mutations linked to AR \cite{alcock2023card}, each varying in prevalence and diagnostic value, identifying the most informative subset is a significant challenge; naive approaches like brute-force testing or selecting the most common AR genes are costly and often suboptimal. 

A method capable of identifying the most informative subset of genes could transform susceptibility testing by targeting the full resistance phenotype, rather than addressing individual resistance types as standard PCR assays do. This would reduce costs and improve efficiency by eliminating redundant measurements, such as co-located resistance genes, and enable laboratories to optimize PCR assays. By consolidating the detection of all resistance types into a single test, this approach simplifies diagnostics while retaining accuracy and practicality for clinical use.

Motivated by the success of reinforcement learning (RL) in solving complex optimization problems, this work explores its application to efficiently identify the most informative gene subsets for PCR and other gene tests. RL's ability to handle high-dimensional state-action spaces while learning complex relationships between different components makes it a good choice for this problem. Predicting resistance against a wide range of antibiotics from a limited subset of gene test results is a partial set classification problem, requiring an approach that can utilize unordered and missing data. Additionally, AR can vary greatly depending on metadata data like country and date, thus these factors need to be accounted for when selecting a subset. Finally, labels in AR datasets are commonly missing due to non-standardized testing procedures, and models therefore need to be able to handle missing labels.

In this study, we address the challenges presented by optimizing gene test selection to enhance AR predictions and our main contributions can be summarized as the following:

\begin{itemize}
    \item We present a model for AR prediction using a novel representation of gene test results.
    \item We introduce a reinforcement learning-based method for training a gene test selection policy that makes use of a trained AR prediction model to evaluate the reward.
    \item We develop two high-performing baselines and conduct extensive evaluations to ensure robust comparison across all methods.
    \item We improve predictive performance by incorporating metadata, demonstrating its significant impact on model accuracy.
\end{itemize}

These contributions are integrated into a comprehensive framework named GenoARM that delivers a tailored gene test selection policy along with a trained antibiotic resistance prediction model. Extensive experimentation on the large-scale National Center for Biotechnology Information (NCBI) \cite{sayers2022ncbi} dataset demonstrates the broad applicability and effectiveness of GenoARM across various pathogens.

\section{Related Work}

\subsection{Deep Learning and Antibiotic Resistance}

Several recent works \cite{chakraborty_deep_2022,kim_machine_2022, ren_prediction_2022}, have identified both the potential and the challenges that exist within the intersection of deep learning and AR prediction, and significant effort is already underway in the area. 

Many methods \cite{tharmakulasingam_explainable_2022, tharmakulasingam_rectified_2023, macesic_predicting_2020, jin_predicting_2024, kuang_accurate_2022} utilize full gene sequences as input data. These methods either utilize dimensionality reduction on the sequences or convert the sequences to genes through existing algorithms. Gene sequences are rich in information, but they require whole-genome sequencing, which is often impractical in clinical routine.

Some attempts \cite{her_pan-genome-based_2018, hyun_machine_2020, kavvas_machine_2018} have been made to identify important AR genes. In  \cite{her_pan-genome-based_2018}, they use a genetic algorithm to select the best performing genes. However, to predict multiple resistances, their method requires a large subset of genes, not achievable by a single PCR test. Others \cite{hyun_machine_2020, kavvas_machine_2018} manage to identify new AR genes along with important existing ones, however, only while utilizing full genome sequences.

\subsection{Feature Selection Methods}

Selecting an optimal subset of genes for testing, based on the predictive performance of a model, can be framed as a black-box optimization problem, where the prediction model serves as the black box. A wide range \cite{sxue_survey_evolutionary_2016} of methods has been proposed to address such problems, including evolutionary algorithms, search-based techniques and heuristic approaches. For instance, evolutionary algorithms have been explored \cite{salimans2017evolutionstrategiesscalablealternative} as alternatives to reinforcement learning for optimizing policy parameters. Several black-box optimization techniques \cite{pudjihartono_review_feature_selection_2022} and heuristic methods \cite{ALIREZANEJAD20201173} have been used to identify key gene features. Although adaptable to our problem, exploring these methods fully is beyond the scope of this study.

\subsection{Active Feature Acquisition}

The challenge of selecting the optimal subset of genes also falls under the umbrella of active feature acquisition (AFA) problems. The majority of research \cite{nam_reinforcement_2021, li_active_2021, fahy_dynamic_2019, yin_reinforcement_2020, shim_joint_2018} in that area considers the problem as a Markov Decision Process (MDP), where at each state, one can either decide to observe another feature or make a prediction. The INVASE framework \cite{yoon_invase_2018} observes all features before selecting the relevant features, which requires all tests to be performed and eliminates the potential gain of only selecting a subset. ODIN \cite{zannone_odin_2019} and many other methods \cite{yu_deep_2023, yin_reinforcement_2020, li_active_2021, nam_reinforcement_2021, shim_joint_2018} assume that once an action has been taken, the new state and the result of that action can be observed (often for a cost). This additional information improves subsequent choices but is not applicable to our problem since doctors would not test genes one at a time but instead test for a set of genes all at once.

\section{Methodology}
\label{sec:method}

\begin{figure}[t!]
\centering
\includegraphics[width=.6\linewidth]{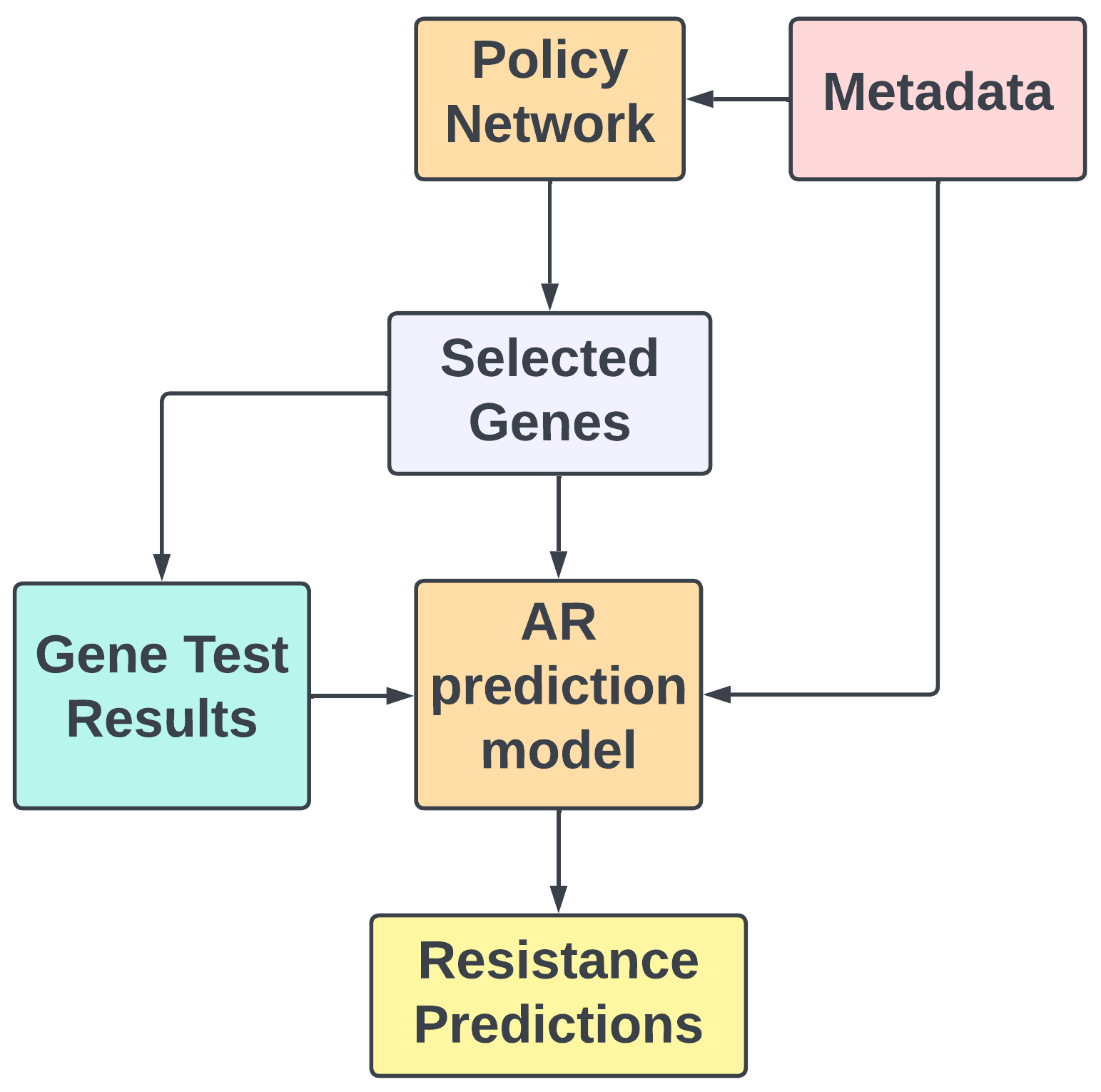}
\caption{An overview of the full GenoARM framework as used during inference.}
\label{fig:full-inference}
\end{figure}

We propose a dual-architecture framework for optimizing PCR test selection in antibiotic resistance prediction, denoted as GenoARM. The framework contains two main modules, an AR prediction model that predicts resistance and a policy network to select the optimal PCR test subset, see Figure ~\ref{fig:full-inference} for more details.

\subsection{Problem Formulation}

The problem of finding an optimal gene test subset can be formulated as follows. Let $G=\{g_1,g_2,\ldots,g_n\}$ be the set of $n$ possible PCR gene tests and denote the result of a test measuring the presence of gene $g_k$ by $e_i(g_k)$ for subject $i$, defined as
\begin{equation}
e_i(g_k) =
\begin{cases}
(1, k), & \text{if the $k$th gene is present}, \\
(0, k), & \text{otherwise}.
\end{cases}
\end{equation}
The test results for a subset of genes $S \subseteq G$ is then given by 
\begin{equation}
e_i(S) = \bigcup_{g \in S} \{e_i(g)\}.
\end{equation}
Let $A = \{ a_1,a_2,\ldots, a_{|A|}\}$ be a set of antibiotics and denote the resistance status of subject $i$ to $A$ as $y_i = (y_i(a_1), y_i(a_2), \ldots, y_i(a_{|A|}))$ where $y_i(a_k) = 0$ if the subject is susceptible, $y_i(a_k) = 1$ if the subject is resistant, and $y_i(a_k) = -1$ if the antibiotic has not been tested. Let $M = \{m_1, m_2, \ldots, m_{|M|}\}$ be a set of unique metadata, and let $o_i \in M$ indicate the metadata observed for subject $i$. Our dataset comes in tuples $\{ e_i(G),y_i, o_i
\}_{i=1}^N$, providing information on gene presence, antibiotic resistance, and metadata for each sample.

The objective is to find subsets $S_m \subset G$ of gene tests for all $ m \in M$ constrained to a maximum size $K$, such that the resistance profile $y_i$ can be accurately predicted for all subjects $i$ with $o_i \in M$. Let $f(e_i(S_{o_i}), o_i)$ denote a function that predicts the probability of resistance of the antibiotics $A$ using the selected subset of gene tests and metadata, and let $t(f(e_i(S_{o_i}), o_i), y_i)$ denote the mean predictive performance of $f$ across all tested antibiotics in $A$. The task is then to maximize the average $t$ across all samples $i$, while limiting the subset size to $K$:
\begin{equation}
\underset{{S_{o_i} \subset G, |S_{o_i}| \le K}}{\max} \quad \frac{1}{N}\sum_{i=1}^N t(f(e_i(S_{o_i}), o_i),y_i).
\end{equation}
Note that each $S_m$, $m\in M $ can be optimized independently. Put simply, we seek a subset of genes, combined with fixed metadata, that maximizes the average predictive performance of the resistance model across all antibiotics.

\subsection{AR Prediction Model}

\begin{figure}[t!]
\centering
\includegraphics[width=0.7\linewidth]{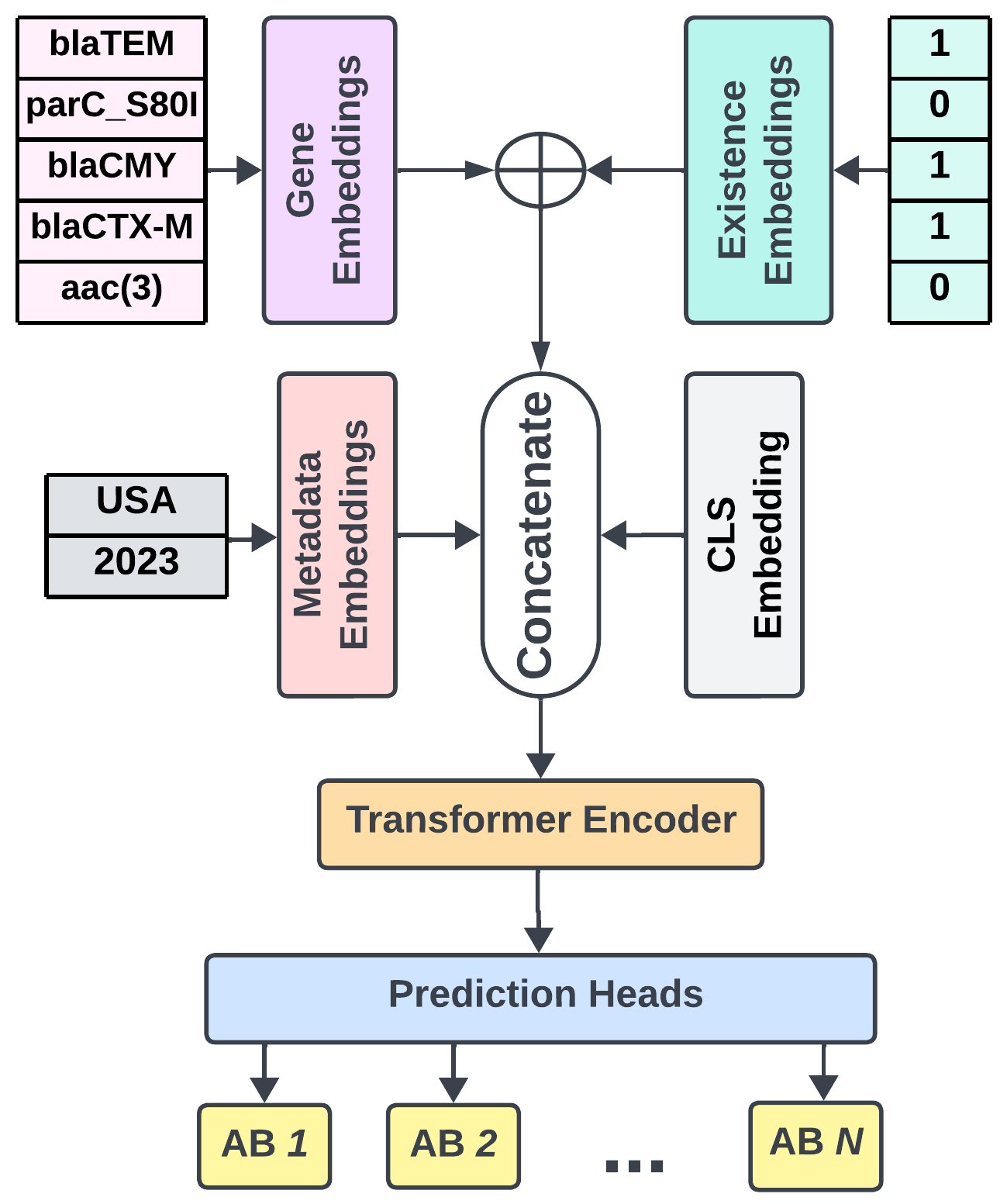}
\caption{Detailed AR prediction model architecture. Gene test names and metadata is shown un-tokenized but would be tokenized prior to being embedded.}
\label{fig:AR-Prediction}
\end{figure}

The AR prediction model's task is to determine the resistance profile of a pathogen sample based on a limited set of results from PCR tests. The results of the PCR test include information about the specific genes tested and the prevalence of those genes within the sample. The output, or resistance profile, is a vector whose elements represent the probability that the sampled bacterium exhibits resistance to different antibiotics. An overview of the architecture can be seen in Figure~\ref{fig:AR-Prediction}.

Formally, we define the AR prediction model as $f_\phi: \{0,1\}^K \times \mathbb{N}^{K}, \mathbb{N}^{|M|} \rightarrow [0,1] ^{|A|}$, which maps $K$ gene features, where $K < |G|$, their corresponding existence maps and the observed metadata to $|A|$ Bernoulli probabilities.

\parsection{Loss function}
The prediction model is parameterized by $\phi$ and trained by directly minimizing the mean binary cross entropy loss over all antibiotics. As antibiotic labels can be missing from the data set, the loss is ignored in those cases.

\parsection{Architecture} The architecture of the AR prediction model contains three main components: First, an embedding module that converts the tokenized gene data, existence maps and observed metadata into a feature vector. Next, a stack of transformer-encoder layers refines these feature vectors \cite{NIPS2017_attention}. Last, $A$ distinct heads, one for each predicted antibiotic. The heads are two-layer multi-layer perceptrons (MLP) using ReLU activations.

The gene tokens and the corresponding existence tokens are processed by separate embedding layers, producing embedding vectors that are combined via addition to create a new feature vector. This method reduces the potential state space from $2*K^{|G|}$ to $K^{|G|} + 2$, significantly simplifying policy optimization. The resulting gene existence vector is then concatenated with the embedded metadata tokens and a CLS token to form the final feature vector. Positional embeddings are excluded, as genes form an unordered set, and their inclusion would hinder learning by increasing input state complexity.

\subsection{Policy Network}
\label{sec:policy_net}

\begin{figure*}[t!]
\centering
\includegraphics[width=0.9\linewidth]{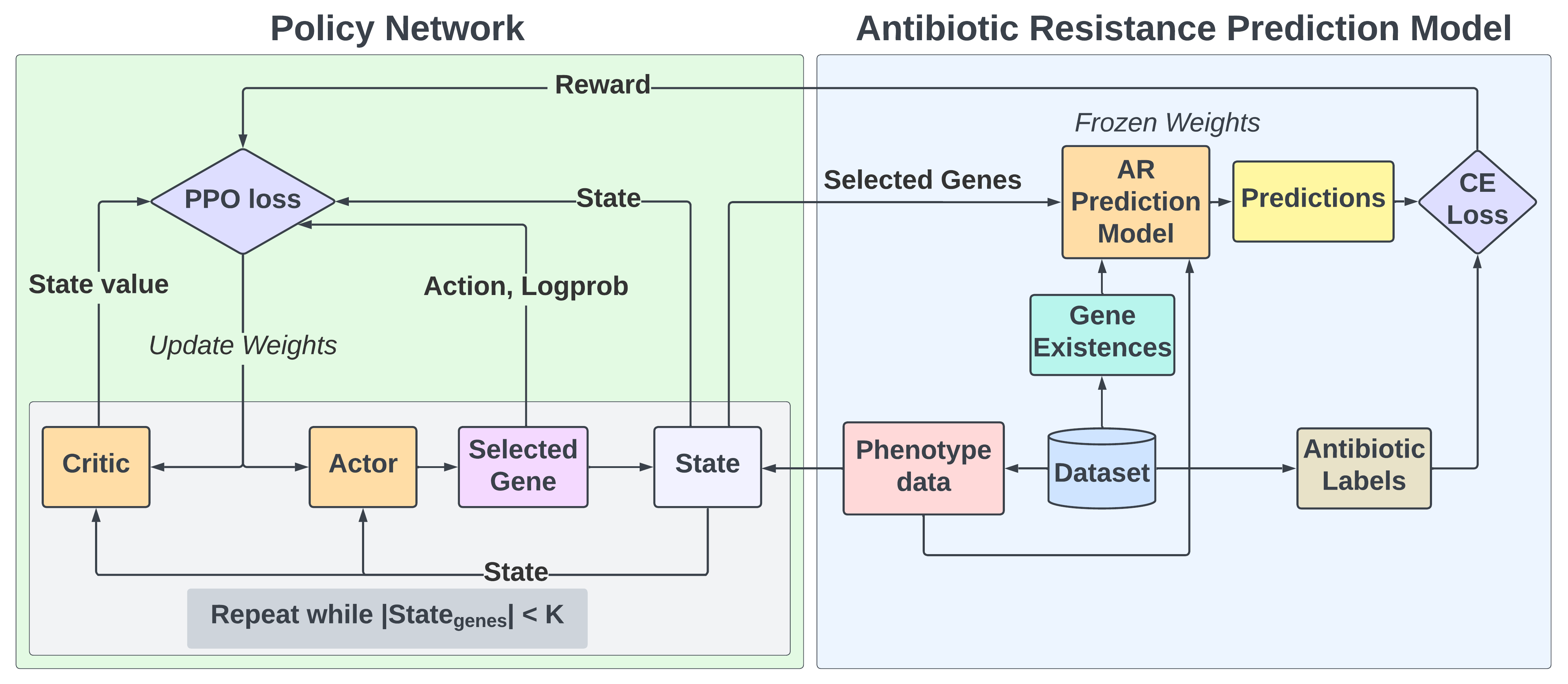}
\caption{An overview of the training framework for the policy network.}
\label{fig:RL-overall}
\end{figure*}

% Describe the problem we are trying to solve
Given a model that predicts the resistance profile of bacteria based on the outcomes of a subset of gene PCR tests, we need the ability to determine which PCR tests to perform. More specifically, we want to know which subset of tests that gives the best predictive performance for the AR prediction model. In more formal terms, we want to find a policy $\pi$ that we can follow to select this subset.

% Explain how/why we instead use a sequential policy.
While the subset can be selected in a single step, we formulate the problem of finding the policy $\pi$ as a sequential problem. Given a set of genes already selected, an optimal policy should then select the subsequent gene to test such that the final subset of genes is optimal once the episode has ended. This allows us to frame it within the structure of an episodic Markov Decision Process (MDP) and utilize reinforcement learning methods. 

% A more formal connection to RL
In this formulation, we seek a policy $\pi(a\vert s)$ in which the policy selects an action based on the current state. The state $s$ contains information on the genes that have already been selected and the action $a$ is defined as selecting a specific gene to test. Once the $K$ genes have been selected, the episode is terminated and the genes and their corresponding test results are given to the AR model. The predictions from the AR model and the ground truth labels are then used to compute the reward ($r$), defined as the negative mean cross-entropy loss over all antibiotics. Since the reward is always zero except at the end of an episode, the return is equal to the reward at the final step with a discount factor of 1.

% The objective of the network.
The objective of the policy network is to maximize the expected return. We optimize the policy through proximal policy optimization \cite{schulman2017proximalpolicyoptimizationalgorithms} and gather data for the policy by allowing it to act in the environment defined by $f_\phi$. An overview of the training framework for the policy network can be found in Figure ~\ref{fig:RL-overall}.

\parsection{Architecture} The policy network is an actor-critic architecture using actor and critic agents with non-shared weights. The actor agent learns the policy $\pi_\phi(a\vert s)$ and is parameterized by $\phi$. The critic agent learns the value function $V_w(s)$ and is parameterized by $w$. Both the actor and the critic are three-layer MLP's using tanh activations. The state $s$ is constructed by a concatenation of a vector $s_g$ representing the currently selected genes and a vector $s_m$ representing the metadata. The vector $s_g$ is created from the sum of currently selected tokens, embedded with a separate learned embedding layer and an MLP. The use of summation ensures that different permutations of the same genes are encoded into an identical state. It reduces the number of possible states down from $\sum_{j=1}^K \frac{N\,!}{(N-j)\,!}$ to $\sum_{j=1}^K \frac{N\,!}{j\,!(N-j)\,!}$ compared to if each different permutation of selected genes was defined as a unique state. The metadata vector $S_m$ is created in a similar way, as a sum of the embedded tokenized metadata. 

\parsection{Choice of reward} While several metrics such as accuracy or Pareto front optimization~\cite{yu_deep_2023}, can be used as rewards, these metric-based rewards tend to be non-smooth. This non-smoothness results in abrupt changes in rewards from minor adjustments in state or input, which can destabilize the learning process. To address this, we utilize the unreduced loss from our prediction model as the reward for selecting a specific set of genes. The unreduced loss serves as an indicator of predictive performance, and provides a more stable and informative reward signal that still aligns with the goal metric.

\parsection{Weight initialization} While the action space for $\pi_\theta$ is large, many genes are not commonly encountered in pathogens, making them inefficient to test, as their results are likely negative in most samples. Our experiments show that finding a good policy in this space can be challenging. To address this, we initialize the last linear layer in the actor module such that the weights are drawn from a normal distribution with a small variance $\sim \mathbf{\cal{N}}(0, 10^{-4})$, while the biases are set to the logarithm of the normalized gene prevalence in the training dataset. This initialization scheme ensures that the initial action distribution aligns with the prevalence of genes, improving convergence during training without restricting the action space.

\subsection{Training Procedure}
\label{sec:train_proc}

The two networks are trained sequentially. First, the AR prediction model is trained using a randomly sampled subset $S_{\texttt{random}}$ of gene tests. The size of $S_{\texttt{random}}$ is set to be larger than the maximum cardinality $K$ but much smaller than the full set of gene tests $|G|$. This ensures that the network is exposed to a wide variety of gene combinations while still focusing on predicting AR from a small subset of genes.

Once the AR prediction model has been trained, its weights are frozen, and the model is used as a static environment. The policy network then interacts with this environment, using its current policy to gather experience. After sufficient experience has been collected, the policy network is updated through proximal policy optimization. This iterative process of gathering experience and updating the policy continues until the policy network has converged.

\section{Results}

In this section, we evaluate the effectiveness of our proposed framework in optimizing PCR test selection for AR prediction. We assess the performance of the reinforcement learning model and the transformer-based AR prediction model across several datasets. To demonstrate the generalizability and robustness of our method it was evaluated on five clinically relevant pathogens: \textit{Klebsiella pneumoniae}, \textit{Escherichia coli}, \textit{Salmonella} , \textit{Staphylococcus aureus}, and \textit{Acinetobacter}.

\subsection{Experimental Setup}

The bulk of our experiments are conducted using a subset of $K = 5$ genes as the most commonly used PCR tests in clinical practice target up to 5 genes. However, there are various types of PCR tests and other targeted methods available for assessing gene prevalence in pathogens. Therefore, we also perform experiments to demonstrate the impact of different values of $K$ on performance. For hyperparameter details we refer to our GitHub repository.

We present results for each species as a weighted sum of evaluation metrics across all antibiotics to account for imbalances in resistance rates and missing labels. The accuracy for each antibiotic is weighted by its number of samples relative to the total number of samples. The very major error and major error are weighted by the number of resistant or susceptible samples relative to the total number of resistant or susceptible samples, respectively.

\subsection{Datasets}

The datasets used in this study are from the publicly available National Center for Biotechnology Information (NCBI) pathogen detection system \cite{sayers2022ncbi}. The splits used for cross-validation is shared on our GitHub repository.

The predicted antibiotics for each species were selected based on specific criteria to ensure reliable analysis. Each antibiotic had to be present in at least one-third of all samples, and the ratio between susceptibility and resistance could not exceed 90\% in either direction. Based on these criteria, the resulting number of predicted antibiotics was 15 for \textit{K.pneumoniae}, 13 for \textit{E.coli}, 10 for \textit{Salmonella}, 8 for \textit{S.aureus}, and 14 for \textit{Acinetobacter}.
The data has been filtered to exclude samples where resistance labels for more than 50\% of the chosen antibiotics are missing. The resistance rates for different antibiotics varied significantly, from 10.5\% resistance for piperacillin-tazobactam to 46\% for ampicillin in the \textit{E.coli} dataset.
The gene test name data has been pre-processed to remove unnecessary annotations that describe specific variants, as many resistance gene classes include sequence variations with the same phenotype, adding no relevant information. For example, ``blaCTX-M-124'' is shortened to ``blaCTX-M''. Not all resistance gene classes have variant-level details, so the pre-processing improves comparability across different types of resistance genes. However, we kept information on amino acid changes in mutated chromosomal genes since they are known to correlate with the induced phenotype.

The metadata used for each sample is the collection date, the isolation source and the geographic location. The collection date contains information regarding both day, month, and year of collection but only the year was used in this study as the focus was to track changes in resistance over longer periods of time. The isolation source, which specifies the origin of the analyzed sample, was pre-processed into broader categories to ensure consistency. For example, ``blood\_whole'' and ``blood\_culture'' were grouped under the general category ``blood''. The geographic location metadata was standardized to reflect the country where each sample was collected.

The resulting size of each dataset after pre-processing is 6111 for \textit{E.coli}, 841 for \textit{K.pneumoniae}, 5961 for \textit{Salmonella}, 1050 for \textit{Acinetobacter}, and 607 for \textit{S.aureus}.

\subsection{Baselines}

Since no existing methods directly address the problem of optimizing PCR test selection for antibiotic resistance prediction, we developed several competitive baselines for comparison. The baselines identify gene test subsets and are evaluated against our proposed frameworks, GenoARM and GenoAR (GenoARM without metadata). 

RandEvolve is a genetic algorithm designed to optimize gene subsets using a trained AR prediction model. The algorithm begins by randomly selecting initial candidate gene subsets based on gene frequencies in the training set. The performance of these candidates is then evaluated using the accuracy of the AR prediction model. The best-performing candidate undergoes mutation, replacing specific genes, and generating new candidates. These mutated candidates are evaluated and compared with the previously best candidate. The best gene tests are then selected and the process is repeated over multiple generations. When metadata is included, RandEvolve finds the best-performing genes for each unique set of metadata values by evaluating candidates on the subset of training data that matches the metadata.

OptStat constructs a gene test subset and a prediction model by iteratively expanding an initially empty subset $V$. For each gene test $g_k \in G$ it forms a new subset $S_k =V \cup \{g_k\}$ and evaluates the prediction accuracy on training data based on the gene test results $e_i(S_k)$. The predictions are determined by selecting the most frequent outcome, resistance or susceptibility, across matching samples in the training data. The subset $S_k$ that yields the highest weighted average accuracy across all antibiotics then replaces $V$. This process repeats until $|V| = K$. When metadata is included, the algorithm generates separate subsets and models for each unique metadata combination, evaluating them on the corresponding training data.

\subsection{Gene Test Optimization, 5 Gene Tests}

We evaluated the performance of GenoARM and all baselines using five-fold cross-validation on the datasets for the five pathogens, optimizing for a subset of $K=5$ gene tests. The results are summarized in ~Table \ref{tab:res_5g_ecoli_kleb_acine} and ~Table \ref{tab:res_5g_salmo_staph}. We also present qualitative example gene subsets from the first fold for \textit{E.coli} in ~Table \ref{tab:ecoli_subset_examples}.

We observe that all evaluated methods have fairly similar results when no metadata is used, with OptStat slightly outperforming the other methods. We also note that four out of five genes were identical for all three evaluated methods, with aph(3b)-I selected in OptStat instead of sul2, improving its predictive performance against aminoglycoside antibiotics. Two of the five selected genes, blaCTX-M and blaTEM, produce resistance against beta-lactam antibiotics. This is a reasonable selection as 6 of the 13 predicted antibiotics are beta-lactam based, including the most difficult, ampicillin, with close to 50\% resistant samples.

The results show that metadata improves performance on all species for models using function approximators to do AR prediction, such as RandEvolve-M and GenoARM, while OptStat-M obtained much smaller increases and in some cases even a decrease in performance. This discrepancy can be attributed to the fact that OptStat-M relies on statistical information derived from each unique metadata value. With fewer available samples for each predictor and gene test policy, the resulting estimates become less reliable, leading to inconsistent performance.

RandEvolve-M also faces challenges due to its gene test policy search being restricted to data associated with the corresponding metadata. That is, to select the gene tests for a certain metadata $m_l$ it only evaluates performance on data with $o_i=m_l$. However, it still performs well, likely due to two key factors. First, the neural network used to perform AR prediction allows RandEvolve-M to generalize predictions effectively, even in scenarios with limited data. Second, despite the large number of unique metadata combinations, the data is heavily clustered around a few common combinations. For instance, in the \textit{E.coli} dataset there are 2550 unique combinations but the 25 most common combinations account for 90\% of the data. The remaining 2525 combinations either have very few or no samples at all. As a result, relatively high performance can be achieved from just these common combinations, which helps to explain RandEvolve-M's robustness despite the metadata constraints.

GenoARM generally achieves the best performance across species when metadata is utilized, though the difference compared to RandEvolve-M is relatively small. The difference could perhaps be attributed to GenoARM’s ability to handle data with less frequent metadata more effectively. Unlike methods that rely on predefined gene test policies, GenoARM’s policy network can evaluate unseen states rather than defaulting to a fixed set of gene tests, since the policy relies on a function approximation.  

In Table~\ref{tab:ecoli_subset_examples} We can observe that the gene test subset found for metadata \{USA, 2020, Urine\} differs from the general subset found by methods not using the metadata. Specifically, the genes aac(3) and sul2 has been replaced by tet(A) and tet(B), both of which confer tetracycline resistance. This change might be explained by the significant increase in tetracycline resistance for urine infections in the USA in 2020, where only ampicillin resistance was higher.

\begin{table*}[!t]
\centering
\begin{tabular}{c||c|c|c||c|c|c||c|c|c}
 & \multicolumn{3}{c||}{\textit{E.coli}} & \multicolumn{3}{c||}{\textit{K.pneumoniae}} & \multicolumn{3}{c}{\textit{Acinetobacter}} \\
\hline
Method & Acc $\uparrow$  & ME $\downarrow$ & VME $\downarrow$ & Acc $\uparrow$  & ME $\downarrow$ & VME $\downarrow$ & Acc $\uparrow$  & ME $\downarrow$ & VME $\downarrow$ \\
\hline
RandEvolve & 89.84 & 0.0395 & 0.2803 & 84.85 & 0.2083 & 0.1203 & 89.23 & 0.1379 & 0.0893 \\
OptStat & \textbf{89.94} & \textbf{0.0357} & 0.2873 & \textbf{85.27} & \textbf{0.2002} & 0.1184 & \textbf{89.80} & \textbf{0.1316} & \textbf{0.0845} \\
GenoAR & 89.91 & 0.0389 & \textbf{0.2797} & 85.16 & 0.2316 & \textbf{0.1181}  & 88.84 & 0.1364 & 0.0962 \\
\hline
\hline
RandEvolve-M & 93.44 & 0.0303 & 0.1670 & 85.70 & 0.2217 & 0.0996 & 90.15 & 0.1423 & 0.0720 \\
OptStat-M & 92.48 & 0.0291 & 0.2078 & 82.59 & 0.2560 & 0.1292 & 87.25 & 0.1950 & 0.0864 \\
GenoARM & \textbf{93.77} & \textbf{0.0260} & \textbf{0.1641} & \textbf{86.80} & \textbf{0.2076} & \textbf{0.0906} & \textbf{90.69} & \textbf{0.1368} & \textbf{0.0665} \\
\end{tabular}
\caption{Antibiotic resistance prediction performance, averaged across all folds, for all methods on \textit{E.coli}, \textit{K.pneumoniae}, and \textit{Acinetobacter} using a subset of 5 gene tests. GenoAR is GenoARM without metadata.}
\label{tab:res_5g_ecoli_kleb_acine}
\end{table*}

\begin{table*}[!t]
\centering
\begin{tabular}{c||c|c|c||c|c|c||c|c|c}
 & \multicolumn{3}{c||}{\textit{Salmonella}} & \multicolumn{3}{c||}{\textit{S.aureus}} & \multicolumn{3}{c}{\textbf{Average}}\\
\hline
Method & Acc $\uparrow$  & ME $\downarrow$ & VME $\downarrow$ & Acc $\uparrow$  & ME $\downarrow$ & VME $\downarrow$ & Acc $\uparrow$  & ME $\downarrow$ & VME $\downarrow$ \\
\hline
RandEvolve & \textbf{92.67} & \textbf{0.0448} & 0.1335 & 91.54 & 0.0340 & 0.2205 & 89.62 & 0.0923 & 0.1688 \\
OptStat & 92.63 & 0.0461 & \textbf{0.1313} & \textbf{91.88} & \textbf{0.0327} & 0.2117 & \textbf{89.90} & \textbf{0.0892} & \textbf{0.1667} \\
GenoAR & 92.48 & 0.0449 & 0.1387 & 91.78 & 0.0352 & \textbf{0.2076} & 89.44 & 0.0974 & 0.1681 \\
\hline
\hline
RandEvolve-M & \textbf{94.54} & 0.0341 & 0.0976 & 91.21 & 0.0474 & 0.1970 & 91.01 & 0.0952 & 0.1267 \\
OptStat-M & 93.23 & \textbf{0.0312} & 0.1444 & \textbf{92.19} & \textbf{0.0379} & 0.1856 & 89.55 & 0.1098 & 0.1507 \\
GenoARM & 94.50 & 0.0375 & \textbf{0.0917} & 91.07 & 0.0578 & \textbf{0.1431} & \textbf{91.36} & \textbf{0.0934} & \textbf{0.1112} \\
\end{tabular}
\caption{Antibiotic resistance prediction performance, averaged across all folds, for all methods on \textit{Salmonella}, \textit{S.aureus}, and the average over all five evaluated pathogens and all folds using a subset of 5 gene tests. GenoAR is GenoARM without metadata.}
\label{tab:res_5g_salmo_staph}
\end{table*}

\begin{table*}[!t]
\centering
\begin{tabular}{c|c|c}
Method & Metadata & Genes  \\
\hline
RandEvolve & - & \{ aac(3), blaCTX-M, blaTEM, parC\_S80I, sul2 \} \\
OptStat & - & \{ aac(3), blaCTX-M, blaTEM, parC\_S80I, aph(3b)-I \} \\
GenoAR & - & \{ aac(3), blaCTX-M, blaTEM, parC\_S80I, sul2 \} \\
RandEvolve-M & \{USA, 2020, Urine\} &  \{blaCTX-M, blaTEM, parC\_S80I, tet(A), tet(B) \} \\
OptStat-M & \{USA, 2020, Urine\} & \{blaCTX-M, blaTEM, parC\_S80I, tet(A), tet(B) \} \\
GenoARM & \{USA, 2020, Urine\} & \{blaCTX-M, blaTEM, parC\_S80I, tet(A), tet(B) \} \\
\end{tabular}
\caption{Example gene test subsets found by all methods on \textit{E.coli} data.}
\label{tab:ecoli_subset_examples}
\end{table*}

\subsection{Ablation Study on Size of Gene Test Subset}

\begin{figure}[t!]
\centering
\includegraphics[width=\linewidth]{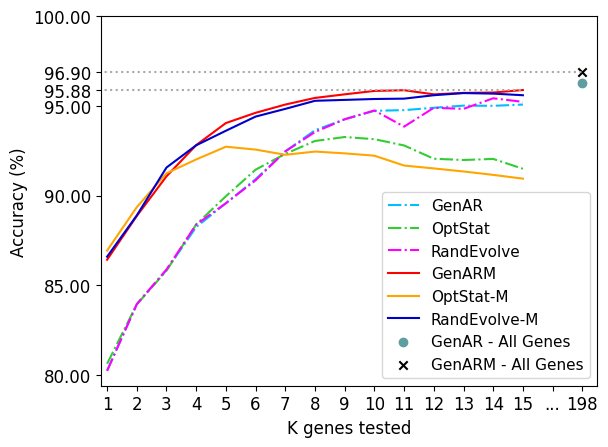}
\caption{Accuracy on \textit{E.coli} of all methods for increasing number of genes. ``GenAR - All genes'' and ``GenARM - All genes'' utilizes all genes in the dataset as inputs.}
\label{fig:one-to-fifteen-genes}
\end{figure}

In Figure ~\ref{fig:one-to-fifteen-genes}, we examine how performance is influenced by the size of the gene test subset $K$. Each model is trained and evaluated separately on the first cross-validation fold for \textit{E.coli} using different values of $K$. Separate AR predictions models have been trained using all AR genes as input. We note that GenAR and RandEvolve perform similarly, with and without metadata. Metadata has a significant impact on performance when $K$ is small, but this effect diminishes as $K$ increases. With a larger gene test budget, the most informative genes can be included, reducing the relative importance of metadata. OptStat's performance peaks at smaller gene subsets and declines for larger $K$ due to missing data for most gene combinations, leading to default predictions.

Finally, we observe that models with an optimized set of 15 genes nearly match the performance of models trained on all 198 genes. This suggests that the number of truly informative genes is relatively small, and that an accurate resistance profile can be predicted from a small set of carefully selected genes.

\subsection{Ablation Study on the Effect of Metadata}

We analyze the impact of each unique type of metadata on performance by comparing models trained with either a single type of metadata, all types of metadata, or no metadata at all on the first cross-validation fold from \textit{E.coli} data. The results, presented in Table~\ref{tab:metadata_effect}, show that each type of metadata enhanced performance compared to the baseline with no metadata, with the most significant improvement coming from the "Source" metadata. Including all types of metadata further enhanced performance, leading to higher accuracy and lower major and very major errors, suggesting that each type of metadata provides potentially complementary information.

\begin{table}[!t]
\centering
\begin{tabular}{c|c|c|c}
 & \multicolumn{3}{c}{\textit{E.coli}}\\
\hline
Method & Acc $\uparrow$  & ME $\downarrow$ & VME $\downarrow$ \\
\hline
GenoAR & 89.57 & 0.0420 & 0.2756 \\
GenoARM (Year) & 91.94 & 0.0338 & 0.2091 \\
GenoARM (Country) & 91.25 & 0.0305 & 0.2438  \\
GenoARM (Source) & 92.33 & 0.0309 & 0.2024  \\
GenoARM (All) & \textbf{94.04} & \textbf{0.0255} & \textbf{0.1531} \\ 
\end{tabular}
\caption{Ablation study over the different choices of metadata used in GenoARM.}
\label{tab:metadata_effect}
\end{table}

\section{Limitations}

This study has several limitations. Our method relies on known AR genes, necessitating retraining to incorporate new genes, which may limit its adaptability to emerging resistance. Additionally, the data used in the study is predominantly from the United States, potentially biasing our models towards this region's AR landscape and limiting generalizability. Including more diverse datasets would improve global applicability.

\section{Conclusion}

In this study, we developed a novel framework that combines reinforcement learning and transformer-based classification models to optimize PCR test selection and improve antibiotic resistance predictions. The optimized PCR tests, along with our high-performing AR prediction model, show significant potential in guiding the selection of appropriate antibiotic treatments. Our results demonstrate that the GenoARM framework performs well across various pathogens effectively utilizing metadata to enhance predictions. The genetic algorithm RandEvolve also proves to be a competitive alternative. Both methods benefit from our proposed AR prediction model, which efficiently utilizes existence embeddings to represent gene presence. Overall, our framework represents a significant advancement in optimizing diagnostic tools for AR in clinical settings. By enhancing the precision of AR predictions and enabling more targeted treatments, our method has the potential to mitigate the impact of AR on global health.

\section{Acknowledgements}

The computations were enabled by resources provided by Chalmers e-Commons at Chalmers and the National Academic Infrastructure for Supercomputing in Sweden (NAISS), partially funded by the Swedish Research Council through grant agreement no. 2022-06725.

\bibliography{main}

\begin{thebibliography}{30}
\providecommand{\natexlab}[1]{#1}

\bibitem[{Alcock et~al.(2023)}]{alcock2023card}
Alcock, B.~P.; et~al. 2023.
\newblock CARD 2023: expanded curation, support for machine learning, and
  resistome prediction at the Comprehensive Antibiotic Resistance Database.
\newblock \emph{Nucleic Acids Research}, 51: D690--D699.

\bibitem[{Alirezanejad et~al.(2020)Alirezanejad, Enayatifar, Motameni, and
  Nematzadeh}]{ALIREZANEJAD20201173}
Alirezanejad, M.; Enayatifar, R.; Motameni, H.; and Nematzadeh, H. 2020.
\newblock Heuristic filter feature selection methods for medical datasets.
\newblock \emph{Genomics}, 112(2): 1173--1181.

\bibitem[{Chakraborty et~al.(2022)Chakraborty, Bhattacharya, Sharma, Roy,
  Islam, Chakraborty, Nandi, and Dhama}]{chakraborty_deep_2022}
Chakraborty, C.; Bhattacharya, M.; Sharma, A.~R.; Roy, S.~S.; Islam, M.~A.;
  Chakraborty, S.; Nandi, S.~S.; and Dhama, K. 2022.
\newblock Deep learning research should be encouraged for diagnosis and
  treatment of antibiotic resistance of microbial infections in treatment
  associated emergencies in hospitals.
\newblock \emph{International Journal of Surgery}, 105.

\bibitem[{Fahy and Yang(2019)}]{fahy_dynamic_2019}
Fahy, C.; and Yang, S. 2019.
\newblock Dynamic {Feature} {Selection} for {Clustering} {High} {Dimensional}
  {Data} {Streams}.
\newblock \emph{IEEE Access}, 7: 127128--127140.
\newblock Conference Name: IEEE Access.

\bibitem[{Her and Wu(2018)}]{her_pan-genome-based_2018}
Her, H.-L.; and Wu, Y.-W. 2018.
\newblock A pan-genome-based machine learning approach for predicting
  antimicrobial resistance activities of the {Escherichia} coli strains.
\newblock \emph{Bioinformatics}, 34(13): i89--i95.

\bibitem[{Hyun et~al.(2020)Hyun, Kavvas, Monk, and Palsson}]{hyun_machine_2020}
Hyun, J.~C.; Kavvas, E.~S.; Monk, J.~M.; and Palsson, B.~O. 2020.
\newblock Machine learning with random subspace ensembles identifies
  antimicrobial resistance determinants from pan-genomes of three pathogens.
\newblock \emph{PLOS Computational Biology}, 16(3): e1007608.
\newblock Publisher: Public Library of Science.

\bibitem[{Jin et~al.(2024)Jin, Jia, Hu, Xu, Shen, and
  Yue}]{jin_predicting_2024}
Jin, C.; Jia, C.; Hu, W.; Xu, H.; Shen, Y.; and Yue, M. 2024.
\newblock Predicting antimicrobial resistance in \textit{{E}. coli} with
  discriminative position fused deep learning classifier.
\newblock \emph{Computational and Structural Biotechnology Journal}, 23:
  559--565.

\bibitem[{Kavvas et~al.(2018)Kavvas, Catoiu, Mih, Yurkovich, Seif, Dillon,
  Heckmann, Anand, Yang, Nizet, Monk, and Palsson}]{kavvas_machine_2018}
Kavvas, E.~S.; Catoiu, E.; Mih, N.; Yurkovich, J.~T.; Seif, Y.; Dillon, N.;
  Heckmann, D.; Anand, A.; Yang, L.; Nizet, V.; Monk, J.~M.; and Palsson, B.~O.
  2018.
\newblock Machine learning and structural analysis of {Mycobacterium}
  tuberculosis pan-genome identifies genetic signatures of antibiotic
  resistance.
\newblock \emph{Nature Communications}, 9: 4306.

\bibitem[{Kim et~al.(2022)Kim, Maguire, Tsang, Gouliouris, Peacock, McAllister,
  McArthur, and Beiko}]{kim_machine_2022}
Kim, J.~I.; Maguire, F.; Tsang, K.~K.; Gouliouris, T.; Peacock, S.~J.;
  McAllister, T.~A.; McArthur, A.~G.; and Beiko, R.~G. 2022.
\newblock Machine {Learning} for {Antimicrobial} {Resistance} {Prediction}:
  {Current} {Practice}, {Limitations}, and {Clinical} {Perspective}.
\newblock \emph{Clinical Microbiology Reviews}, 35(3): e00179--21.
\newblock Publisher: American Society for Microbiology.

\bibitem[{Kuang et~al.(2022)Kuang, Wang, Hernandez, Zhang, and
  Grossman}]{kuang_accurate_2022}
Kuang, X.; Wang, F.; Hernandez, K.~M.; Zhang, Z.; and Grossman, R.~L. 2022.
\newblock Accurate and rapid prediction of tuberculosis drug resistance from
  genome sequence data using traditional machine learning algorithms and {CNN}.
\newblock \emph{Scientific Reports}, 12(1): 2427.
\newblock Number: 1 Publisher: Nature Publishing Group.

\bibitem[{Laxminarayan et~al.(2013)Laxminarayan, Duse, Wattal, Zaidi, Wertheim,
  Sumpradit, Vlieghe, Hara, Gould, Goossens
  et~al.}]{laxminarayan2013antibiotic}
Laxminarayan, R.; Duse, A.; Wattal, C.; Zaidi, A.~K.; Wertheim, H.~F.;
  Sumpradit, N.; Vlieghe, E.; Hara, G.~L.; Gould, I.~M.; Goossens, H.; et~al.
  2013.
\newblock Antibiotic resistance—the need for global solutions.
\newblock \emph{The Lancet infectious diseases}, 13(12): 1057--1098.

\bibitem[{Li and Oliva(2021)}]{li_active_2021}
Li, Y.; and Oliva, J. 2021.
\newblock Active {Feature} {Acquisition} with {Generative} {Surrogate}
  {Models}.
\newblock In \emph{Proceedings of the 38th {International} {Conference} on
  {Machine} {Learning}}, 6450--6459. PMLR.
\newblock ISSN: 2640-3498.

\bibitem[{Macesic et~al.(2020)Macesic, Bear Don’t~Walk, Pe’er, Tatonetti,
  Peleg, and Uhlemann}]{macesic_predicting_2020}
Macesic, N.; Bear Don’t~Walk, O.~J.; Pe’er, I.; Tatonetti, N.~P.; Peleg,
  A.~Y.; and Uhlemann, A.-C. 2020.
\newblock Predicting {Phenotypic} {Polymyxin} {Resistance} in {Klebsiella}
  pneumoniae through {Machine} {Learning} {Analysis} of {Genomic} {Data}.
\newblock \emph{mSystems}, 5(3): 10.1128/msystems.00656--19.
\newblock Publisher: American Society for Microbiology.

\bibitem[{Nam, Fleming, and Brunskill(2021)}]{nam_reinforcement_2021}
Nam, H.~A.; Fleming, S.; and Brunskill, E. 2021.
\newblock Reinforcement {Learning} with {State} {Observation} {Costs} in
  {Action}-{Contingent} {Noiselessly} {Observable} {Markov} {Decision}
  {Processes}.
\newblock In \emph{Advances in {Neural} {Information} {Processing} {Systems}},
  volume~34, 15650--15666. Curran Associates, Inc.

\bibitem[{O'Neill(2014)}]{o2014antimicrobial}
O'Neill, J. 2014.
\newblock Antimicrobial resistance: tackling a crisis for the health and wealth
  of nations.
\newblock \emph{Rev. Antimicrob. Resist.}

\bibitem[{O'Neill(2016)}]{o2016tackling}
O'Neill, J. 2016.
\newblock Tackling drug-resistant infections globally: final report and
  recommendations.

\bibitem[{Pudjihartono et~al.(2022)Pudjihartono, Fadason, Kempa-Liehr, and
  O'Sullivan}]{pudjihartono_review_feature_selection_2022}
Pudjihartono, N.; Fadason, T.; Kempa-Liehr, A.~W.; and O'Sullivan, J.~M. 2022.
\newblock A Review of Feature Selection Methods for Machine Learning-Based
  Disease Risk Prediction.
\newblock \emph{Frontiers in bioinformatics}, 2.

\bibitem[{Ren et~al.(2022)Ren, Chakraborty, Doijad, Falgenhauer, Falgenhauer,
  Goesmann, Hauschild, Schwengers, and Heider}]{ren_prediction_2022}
Ren, Y.; Chakraborty, T.; Doijad, S.; Falgenhauer, L.; Falgenhauer, J.;
  Goesmann, A.; Hauschild, A.-C.; Schwengers, O.; and Heider, D. 2022.
\newblock Prediction of antimicrobial resistance based on whole-genome
  sequencing and machine learning.
\newblock \emph{Bioinformatics}, 38(2): 325--334.

\bibitem[{Salimans et~al.(2017)Salimans, Ho, Chen, Sidor, and
  Sutskever}]{salimans2017evolutionstrategiesscalablealternative}
Salimans, T.; Ho, J.; Chen, X.; Sidor, S.; and Sutskever, I. 2017.
\newblock Evolution Strategies as a Scalable Alternative to Reinforcement
  Learning.
\newblock arXiv:1703.03864.

\bibitem[{Sayers et~al.(2022)}]{sayers2022ncbi}
Sayers, E., W; et~al. 2022.
\newblock Database resources of the national center for biotechnology
  information.
\newblock \emph{Nucleic acids research}, 50(D1).

\bibitem[{Schulman et~al.(2017)Schulman, Wolski, Dhariwal, Radford, and
  Klimov}]{schulman2017proximalpolicyoptimizationalgorithms}
Schulman, J.; Wolski, F.; Dhariwal, P.; Radford, A.; and Klimov, O. 2017.
\newblock Proximal Policy Optimization Algorithms.
\newblock arXiv:1707.06347.

\bibitem[{Shim, Hwang, and Yang(2018)}]{shim_joint_2018}
Shim, H.; Hwang, S.~J.; and Yang, E. 2018.
\newblock Joint {Active} {Feature} {Acquisition} and {Classification} with
  {Variable}-{Size} {Set} {Encoding}.
\newblock In \emph{Advances in {Neural} {Information} {Processing} {Systems}},
  volume~31. Curran Associates, Inc.

\bibitem[{Tharmakulasingam et~al.(2023)Tharmakulasingam, Gardner, La~Ragione,
  and Fernando}]{tharmakulasingam_rectified_2023}
Tharmakulasingam, M.; Gardner, B.; La~Ragione, R.; and Fernando, A. 2023.
\newblock Rectified {Classifier} {Chains} for {Prediction} of {Antibiotic}
  {Resistance} {From} {Multi}-{Labelled} {Data} {With} {Missing} {Labels}.
\newblock \emph{IEEE/ACM Transactions on Computational Biology and
  Bioinformatics}, 20(1): 625--636.
\newblock Conference Name: IEEE/ACM Transactions on Computational Biology and
  Bioinformatics.

\bibitem[{Tharmakulasingam et~al.(2022)Tharmakulasingam, Gardner, Ragione, and
  Fernando}]{tharmakulasingam_explainable_2022}
Tharmakulasingam, M.; Gardner, B.; Ragione, R.~L.; and Fernando, A. 2022.
\newblock Explainable {Deep} {Learning} {Approach} for {Multilabel}
  {Classification} of {Antimicrobial} {Resistance} {With} {Missing} {Labels}.
\newblock \emph{IEEE Access}, 10: 113073--113085.
\newblock Conference Name: IEEE Access.

\bibitem[{Vaswani et~al.(2017)Vaswani, Shazeer, Parmar, Uszkoreit, Jones,
  Gomez, Kaiser, and Polosukhin}]{NIPS2017_attention}
Vaswani, A.; Shazeer, N.; Parmar, N.; Uszkoreit, J.; Jones, L.; Gomez, A.~N.;
  Kaiser, L.~u.; and Polosukhin, I. 2017.
\newblock Attention is All you Need.
\newblock In Guyon, I.; Luxburg, U.~V.; Bengio, S.; Wallach, H.; Fergus, R.;
  Vishwanathan, S.; and Garnett, R., eds., \emph{Advances in Neural Information
  Processing Systems}, volume~30. Curran Associates, Inc.

\bibitem[{Xue et~al.(2016)Xue, Zhang, Browne, and
  Yao}]{sxue_survey_evolutionary_2016}
Xue, B.; Zhang, M.; Browne, W.~N.; and Yao, X. 2016.
\newblock A Survey on Evolutionary Computation Approaches to Feature Selection.
\newblock \emph{IEEE Transactions on Evolutionary Computation}, 20(4):
  606--626.

\bibitem[{Yin et~al.(2020)Yin, Li, Pan, Zhang, and
  Tschiatschek}]{yin_reinforcement_2020}
Yin, H.; Li, Y.; Pan, S.~J.; Zhang, C.; and Tschiatschek, S. 2020.
\newblock Reinforcement {Learning} with {Efficient} {Active} {Feature}
  {Acquisition}.
\newblock ArXiv:2011.00825 [cs].

\bibitem[{Yoon, Jordon, and Schaar(2018)}]{yoon_invase_2018}
Yoon, J.; Jordon, J.; and Schaar, M. v.~d. 2018.
\newblock {INVASE}: {Instance}-wise {Variable} {Selection} using {Neural}
  {Networks}.
\newblock In \emph{International Conference on Learning Representations}.

\bibitem[{Yu et~al.(2023)Yu, Li, Kim, Huang, Luo, and Wang}]{yu_deep_2023}
Yu, Z.; Li, Y.; Kim, J.; Huang, K.; Luo, Y.; and Wang, M. 2023.
\newblock Deep {Reinforcement} {Learning} for {Cost}-{Effective} {Medical}
  {Diagnosis}.
\newblock ArXiv:2302.10261 [cs].

\bibitem[{Zannone et~al.(2019)Zannone, Lobato, Zhang, and
  Palla}]{zannone_odin_2019}
Zannone, S.; Lobato, J. M.~H.; Zhang, C.; and Palla, K. 2019.
\newblock {ODIN}: {Optimal} {Discovery} of {High}-value {INformation} {Using}
  {Model}-based {Deep} {Reinforcement} {Learning}.
\newblock In \emph{Real-world Sequential Decision Making Workshop, ICML}.

\end{thebibliography}
\end{document}